\documentstyle[12pt, epsfig]{article}
\begin{document}

\vspace*{2.0cm}
\noindent{\large Probing quantum statistical mechanics with Bose gases:\\ 
Non-trivial order parameter topology from a Bose-Einstein quench}

\vspace{0.5cm}

\noindent{J.R.
Anglin$^{a,b}$
and W.H. Zurek$^b$
\vspace{0.5cm}

\noindent $^a$Institut f\"ur Theoretische Physik,
Universit\"at Innsbruck, Austria

\vspace{0.5cm}

\noindent$^b$T-6, MS B288, Los
Alamos National Laboratory, Los Alamos, New Mexico 87545

\vspace{0.75cm}
                                                  
A rapid second order phase transition 
can have a significant probability of producing a metastable state instead of
the equilibrium state.  We consider the case of rapid Bose-Einstein condensation in a 
toroidal trap resulting in a spontaneous superfluid current, and
compare the phenomenological time-dependent Ginzburg-Landau theory with quantum kinetic
theory.  A simple model suggests the effect should be observable.

\vspace{0.75cm}

\noindent{\bf 1. THE ORDER PARAMETER AND THE EFFECTIVE POTENTIAL}\\

{\parfillskip=0pt 
A qualitative understanding of a second order phase transition may be had by considering
the system to be described by a two-component {\it order parameter}, consisting of a
modulus $R$ and an angle $\theta$[1].  The order parameter is taken to behave as a
particle in a two-dimensional effective potential $V(R)$, with some form of dissipation
dragging the system towards the bottom of this potential (possibly opposed by some random
noise).}

\begin{center}\parbox[c]{0.8\linewidth}{
\epsfig{file=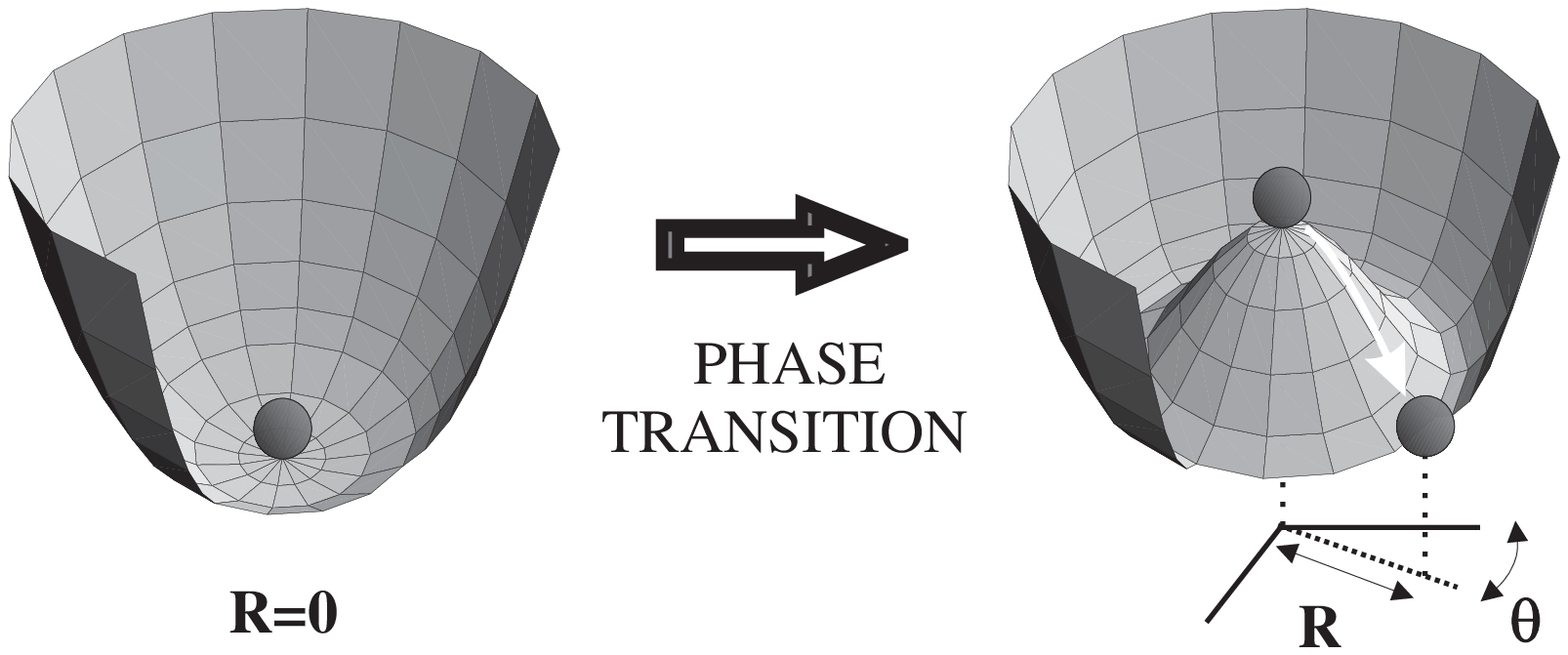,width=\linewidth}\begin{center} 
Figure 1: Effective potential theory of a second order transition.\end{center}}
\end{center}  
{\noindent}In the disordered phase of the system, the minimum of $V(R)$ is at $R=0$,
so that $\theta$ is indefinite (or purely random) (Fig.~1, left picture).  
The phase transition
then occurs through a modification of the effective potential, such that $R=0$ becomes a
local maximum, and the new minimum is a circular trough at some finite radius.  The system
rolls down the central hill to the trough, (Fig.~1, right picture) and $\theta$ 
acquires a definite value (with only small fluctuations due to noise).  Which definite value 
$\theta$ acquires, however, is random, determined by small fluctuations while the system is 
near $R=0$.

One can also allow $R,\theta$ to be functions of position (with $V(R)$
assumed to have the same form everywhere).  In the ordered phase, the definite angle
$\theta(x)$ may 
then vary spatially.  A qualitative question thus arises, namely the topology of
 $\theta(x)$.  Suppose we let $x$
lie on a circle, for example: 
what {\it winding number} $\oint\!dx\,\partial_x\theta /(2\pi)$ should we 
typically observe after a transition?  If the transition is slow, $\theta(x)$ will have time
to organize itself all around the circle into some lowest energy configuration (such as
$\theta(x)$ constant); but if it is fast, the random choice of $\theta$ around the circle
will have some finite correlation length, and non-zero
winding number will sometimes result.  If it does, it will tend to be stable, since there is 
a high effective potential barrier for $R=0$, and without making 
$R(x)\to 0$ somewhere, it is impossible to alter the winding number.  See Fig.~2.\\

\begin{center}\parbox[c]{.8\linewidth}{
\epsfig{file=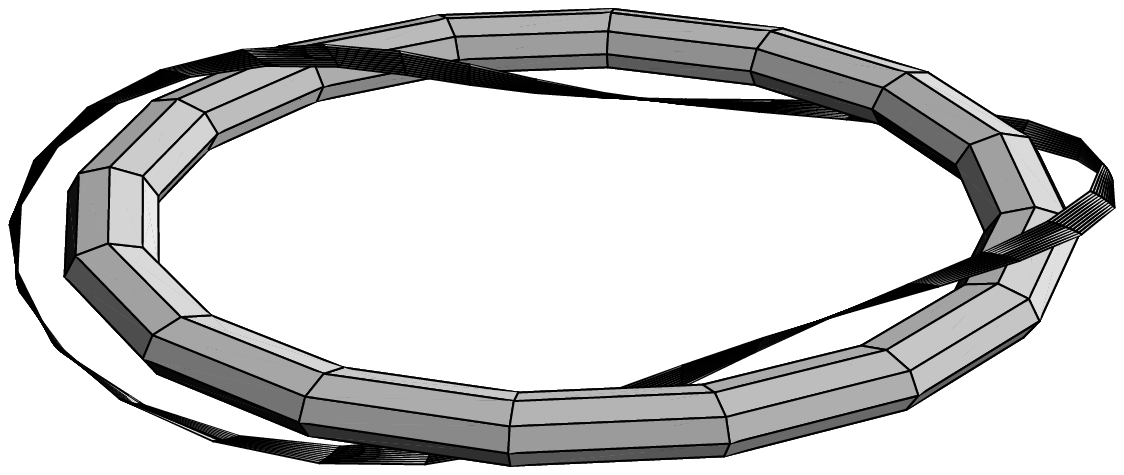,width=\linewidth}
\begin{center}Figure 2: Winding number two; distance of ribbon from torus represents $R(x)$,
angle of winding around torus represents $\theta(x)$.\end{center}}\end{center} 

The emergence of non-trivial topology from a rapid (´quench-like´) phase 
transition is common to a wide range of physical systems, real and postulated, from vortex
loops in superfluid helium[2,3] to cosmic strings in the early universe[4,2].  
With the advent of dilute alkali gas Bose condensates[5], we can now 
investigate this basic statistical mechanical problem in a new arena.  
The condensate mean field wave function
$\psi = Re^{i\theta}$ provides our order parameter; and the current experimental technique 
of evaporative cooling automatically provides a rapid phase transition (on the Boltzmann
scattering time scale).  Our example of $x$ lying on a circle can realistically be
achieved with a toroidal trap, sufficiently tight
for the system to be approximately one-dimensional.  (We will consider such a trap
of circumference ${\sim}100\mu$m, which we understand is a 
credible prospect for the relatively near future.)  In this context, 
non-zero winding number implies circulation around the trap, since 
$\nabla \theta$ 
is the superfluid velocity.  (Angular momentum conservation is an issue we will 
discuss briefly below, only noting for now that the condensate is not isolated.)\\

\noindent{\bf 2. TIME-DEPENDENT GINZBURG-LANDAU THEORY}\\

To proceed to a quantitative description of our subject, we first consider the
phenomenological theory obtained by letting the time evolution of $\psi$ be governed
by a first order equation involving a potential of the Ginzburg-Landau form we have sketched 
above:
\begin{equation}\label{TDGL} 
\tau_0\dot\psi = \beta\Bigl({\hbar^2\over2M}\nabla^2 +\mu -\Lambda 
|\psi|^2\Bigr)\psi\;, 
\end{equation} 
where $\beta=(k_B T)^{-1}$, and $\tau_0$ and $\Lambda>0$ are 
phenomenological parameters.  The thermodynamical variable $\mu$ behaves 
near the critical point, in the case we consider, as
\begin{equation}\label{muTc}
\mu = {3\over2}(T_c -T) + {\cal O}(T_c-T)^2\;,
\end{equation}
where $T_c$ is the critical temperature.  The equilibration time for long 
wavelengths is $\tau=\tau_0 k_BT/|\mu|$.  The system's
disordered phase is described by $\mu<0$; the ordered phase appears when $\mu>0$.
Note that this time-dependent Ginzburg-Landau (TDGL) theory also typically assumes some 
small stochastic forces acting on $\psi$; we will leave these implicit.

A quench occurs if $\mu$ changes with time from negative to positive values.
The divergence of the equilibration time $\tau$ at the critical point $\mu=0$ is
associated with {\it critical slowing down}.  Because of this critical slowing
down, ${d\mu\over dt}/\mu$ must exceed $1/\tau$ in some neighbourhood of the 
critical
point, and so there must be an epoch in which the system is out of equilibrium.
What are at the beginning of this epoch mere fluctuations in the disordered
phase, in which higher energy modes happen momentarily to be more populated than
the lowest mode, can thus pass unsuppressed by equilibration into the ordered
phase, to become topologically non-trivial configurations of $\psi(x)$.  

The interval within which equilibration is negligible can be identified as the
period wherein $|t|/\tau <1$.  If we define the quench time scale $\tau_Q$ by
letting $\beta\mu = t/\tau_Q$ (choosing $t=0$ as the moment the system crosses
the critical point), this implies that the crucial interval is
$-\hat{t}<t<\hat{t}$, for $\hat{t} = \sqrt{\tau_Q\tau_0}$.  The
correlation length $\hat{\xi}$ for fluctuations at time $t=-\hat{t}$ is then
given by $\hbar/(2M\hat{\xi}^2)=\mu(-\hat{t})$, which (assuming
$T(-\hat{t})\doteq T_c$) implies that $\hat{\xi}=\lambda_{T_c}
(\tau_Q/\tau_0)^{1/4}$, for $\lambda_{T}=\hbar(2Mk_BT)^{-{1\over2}}$ the thermal
de Broglie wavelength[2].  Assuming
one independently chosen phase within each correlation length $\hat \xi$, then
modeling the phase distribution around the torus as a random walk suggests that
the net winding number should be of order $\sqrt{L/\hat{\xi}}$ for $L$ the trap 
circumference.  Since evaporative
cooling may be expected to yield $(\tau_Q/\tau_0)^{1/4}$ of order one, and $T_c$ is
several hundred nK, we can estimate $\hat{\xi}\sim 100$ nm, 
leading us to expect winding numbers in our $100 \mu$m torus of order ten.  At current
experimental densities, this implies a current approaching the Landau critical velocity,
in the range of mm/s.      

Considering this intriguing possibility raises an
obvious question:  is TDGL actually relevant to finite samples of dilute gas,
far from equilibrium?\\  

\noindent{\bf 3. QUANTUM KINETIC THEORY}\\

Because the condensates now available are dilute enough to be weakly interacting, there are
good prospects for answering this question theoretically, by constructing from first 
principles
a quantum kinetic theory to describe the whole process accurately.  Using a second-quantized
description of the trapped gas, one treats all modes above some judiciously chosen energy 
level as a reservoir of
particles, coupled to the lower modes by two-particle scattering.  Tracing over the
reservoir modes leads to a 
master equation for the low energy modes[6], very similar to the
equation for a multimode laser.  Scattering of particles from the reservoir modes into the low
modes, and vice versa, provides gain and loss terms.  If the reservoir is 
described at all times by a grand canonical ensemble (of time-dependent temperature and 
chemical potential, in general), then the gain and loss processes are related by a type of 
fluctuation-dissipation relation, in which the reservoir chemical potential and the 
repulsive self-interaction of the gas combine in precisely the Ginzburg-Landau form.  As a 
result, as long as these gain and loss processes are the only significant scattering channels,
and the condensate self-Hamiltonian can be linearized to a good approximation, the system is 
indeed described quite well by time-dependent Ginzburg-Landau theory.  Since these two
conditions can be shown to hold near the critical point, all
the results of section 2 can be recovered from quantum kinetic theory.

We can also clarify how the onset of a runaway
process like Bose-Einstein condensation by evaporation can be consistent with critical 
slowing down: though the particle numbers in the lowest modes of the trapped gas are 
growing very fast, the required particle numbers to be in equilibrium with the higher energy
modes grow much faster still as the temperature passes through $T_c$.  So even with 
Bose-enhanced scattering at ever increasing rates, the rate at which the system is 
able to maintain equilibrium actually decreases.

The problem is that circulating states 
only become metastable above a threshold density, precisely because nonlinearity is
required.  So understanding the probability for a circulating condensate to grow into
metastability requires extending quantum kinetic theory into the fully nonlinear regime,
and without begging the basic question by assuming from the start that some particular
mode will end up with the condensate.\\  

\noindent{\bf 3.1. A two-mode model}

As a first step towards this goal, we consider a
toy model of just two competing modes, such as the torus modes of different winding numbers 
$k_0$, $k_1$.
The self-Hamiltonian of this system is taken as\\
 
\noindent\parbox[l]{\linewidth}
{${\displaystyle\hat{H} = E [\hat{n}_1 + {1\over 2N_c} (\hat{n}_0^2 + \hat{n}_1^2 + 4 
\hat{n}_1\hat{n}_0)]\;.}$\hfill (3)}\\

\noindent
Since the self-Hamiltonian must conserve both particle number and angular momentum, it
can only be a function of $\hat n_0$ and $\hat n_1$ (which greatly simplifies the derivation
of the master equation).
Because we have incorporated the Bose enhancement of inter-mode
repulsion (the factor of 4 instead of 2 in front of the
$\hat{n}_0\hat{n}_1$ term, which is of course the best case value,
obtained when the two orbitals overlap completely), we make the state
with all particles in the 1 mode a local minimum of the energy for fixed $n_0
+ n_1 > (N_c+1)$.  We have thus obtained an simple model which
exhibits metastability above a threshold.

To make our results more meaningful, we estimate the experimental ranges of our
parameters $E$ and $N_c$.
For our toroidal trap $\beta E$ would be on the order of $10^{-4}$
at current experimental temperatures, and proportional to $k_1^2-k_0^2$.  
$N_c$ would be of order one for $(k_0,k_1)=(0,1)$, but higher for higher
modes (since $E/N_c$ is actually constant).  

\begin{center}\parbox[c]{.8\linewidth}{
\epsfig{file=wbqfix2.EPS,width=\linewidth}
Figure 3: 
Trajectories from QKT (solid) and TDGL (dotted); heavy dashed line is 
threshold for metastability of 
mode 1.  Initial times are $\hat{t}$; quench is $\beta\mu=\tanh(t/\tau_Q)$, 
$\beta=\beta_c e^{\tanh(t/\tau_Q)}$.  
Parameters are $N_c=100$, and (a) $\Gamma\tau_Q=10$, $\beta_c E=0.01$; (b) 
$\Gamma\tau_Q = 100$, $\beta_c E = 0.05$.  All trajectories shown initially 
have $n_1>n_0$.}\end{center}

Tracing out a reservoir of higher modes as described above
leads to a master equation for the two 
modes 0 and 1, which even for this highly simplified model is only directly tractable 
numerically.  Here we will simply consider it as a kind of Fokker-Plank equation for 
the particle numbers $n_0,n_1$,
and by neglecting the diffusion terms (valid for fast quenches after the early stage,
which can be solved separately by assuming linearity),
we extract an equation of motion for $n_0$ and $n_1$ which we can compare to TDGL:
\begin{eqnarray}\label{traj} 
\dot{n}_0 &=& \Gamma  n_0
\Bigl[e^{\beta\mu} -e^{{\beta E\over N_c}(n_0+2n_1)}
+2\beta E n_1 e^{-{\beta E\over2N_c}|N_c+n_0-n_1|} 
	\sinh{\beta E\over2N_c}(N_c+n_0-n_1)\Bigr]\nonumber\\ 
\dot{n}_1 &=& \Gamma n_1\Bigl[e^{\beta\mu} 
	-e^{{\beta E\over N_c}(N_c+n_1+2n_0)}
	-2\beta E n_0 e^{-{\beta E\over2N_c}|N_c+n_0-n_1|} 
	\sinh{\beta E\over2N_c}(N_c+n_0-n_1)\Bigr]\nonumber .  
\end{eqnarray}

When
$n_0+2n_1$ and $N_c+n_1+2n_0$ are both close to $N_c\mu /E$, or for
low enough particle numbers, we may replace $n_j\to|\psi_j|^2$ in the
first two terms in each equation of (\ref{traj}) to obtain a TDGL equation,
in the sense that $\dot\psi_j$ is set equal to the variation of a
Ginzburg-Landau effective potential with respect to $\psi_j^*$.  But
the last term in each equation is not of Ginzburg-Landau form (it
does not even involve $\mu$).  These non-GL terms conserve $n_0+n_1$,
and so are neither gain nor loss but another scattering channel
hitherto neglected: collisions with
reservoir particles shifting particles between our two low modes.  
These processes are insignificant at low particle numbers, but grow much stronger as
the condensate grows larger, because they are doubly Bose-enhanced.  And they turn out 
to have a simple but potentially drastic effect:
they allow they system to equilibrate in energy faster than in particle number.

Representative solutions to (\ref{traj}) are shown in Fig.~3, together
with the $|\psi_j|^2$ given by the TDGL theory formed by keeping only
the first two terms of each equation in (\ref{traj}), expanding the
exponentials to first order.  It is clear that for sufficiently
fast quenches the two theories accord quite well, but that for slower
quenches TDGL significantly overestimates the probability of reaching
the metastable state, because more rapid equilibration in energy than particle
number strongly favours the
lowest mode.  This is the main new feature we have identified in dilute Bose
gases as a GL-like system.\\

\noindent{\bf 3.2. Angular momentum}
 
Finally, we return to the question of angular momentum.  
Spontaneous appearance of a circulating state, condensing out of a cold atomic
cloud with no net rotation, obviously implies angular momentum concentration in
the condensate.  Compensating ejection of fast atoms with opposite angular momentum during
the evaporation process, in which after all most of the initial atoms are expelled, is a 
plausible mechanism by which this can proceed.  Our assumption
that the non-condensate atoms remain in a grand canonical ensemble relies implicitly on
some such process.  In our two-mode model, constraining the reservoir's mean angular momentum 
to be always opposite to that of the condensate will effectively just raise $E$ and $N_c$.
Angular momentum transport during evaporative cooling has as yet 
received no detailed study, however.\\

\noindent{\bf 4. CONCLUSIONS}\\

Despite the shortcomings of TDGL revealed by our toy model, we emphasize that
kinetic theory does show that TDGL is indeed relevant to
trapped dilute gases: what TDGL
requires is not outright rejection, but corrections.
And although the corrections may be substantial,
the qualitative features predicted by TDGL are still recovered.  
While the extension of quantum kinetic
theory beyond toy models, to realistic descriptions of spontaneous currents, 
will obviously require further study, the prospects 
for experimental realization of spontaneous
currents are quite
encouraging.  Athough in this brief treatment we have neglected noise, 
diffusive nucleation actually 
gives a lower bound of typical winding number one for condensation
in our hundred micron toroidal trap; with $10^6$ atoms, even this
has an equilibrium probability of order $e^{-10}$.

\vspace*{.75cm}
\noindent{\bf REFERENCES}

{\begin{list}{}
{\setlength{\leftmargin}{1.5em}
\setlength{\parindent}{0pt} 
\setlength{\labelwidth}{1em}
\setlength{\itemsep}{0pt}
\setlength{\parsep}{0pt}}
\item[1.] V.L.~Ginzburg and L.D.~Landau, Zh. Eksperim. i. Teor. Fiz. {\bf 20}, 1064 (1950)	.
\item[2.] W.H.~Zurek, Nature {\bf 317}, 505 (1985); Acta Physica Polonica 
{\bf B 24}, 1301 (1993); Phys. Rep. {\bf 276}, 177 (1996).
\item[3.] P.C.~Hendry {\it et al.}, Nature {\bf 368}, 315 (1994); 
V.M.H.~Ruutu {\it et al.}, Nature {\bf 382}, 332 (1996); C.~Ba\"uerle {\it et 
al.}, Nature {\bf 382}, 334 (1996).
\item[4.] T.W.B.~Kibble, J. Phys.  {\bf A 9}, 1387 (1976).
\item[5.] M. Anderson {\it et al.}, Science {\bf 269}, 198 (1995); 
K.B.~Davis {\it et al.}, Phys. Rev. Lett. {\bf 75}, 3969 (1995); C.C.~Bradley 
{\it et al.}, Phys. Rev. Lett. {\bf 75}, 1687 (1995).
\item[6.] C.W.~Gardiner and P.~Zoller, Phys. Rev. {\bf A55}, 2901 (1997); 
cond-mat/9712002; D.~Jaksch, C.W.~Gardiner, and P. Zoller, Phys. Rev. {\bf A56}, 
575 (1997); J.R.~Anglin, Phys. Rev. Lett. {\bf 79}, 6 (1997). 
\end{list}}

\end{document}